\documentclass[rmp, aps, twocolumn, longbibliography]{revtex4-1}
\usepackage{epsfig}
\usepackage{amssymb}
\usepackage{epstopdf}
\usepackage{amsmath}
\usepackage{bm}
\usepackage{xcolor}
\usepackage{soul}

\begin{document}
\count\footins = 1000

\title{Quantum Computation and Arrows of Time}
\author{Nathan Argaman}
\email{argaman@mailaps.org}
\affiliation{Department of Physics, Nuclear Research Center -- Negev, P.O. Box 9001, Be'er Sheva 84190, Israel}

\begin{abstract}
Quantum physics is surprising in many ways.  One surprise is the threat to locality implied by Bell's Theorem.  Another surprise is the capacity of quantum computation, which poses a threat to the complexity-theoretic Church-Turing thesis.  In both cases, the surprise may be due to taking for granted a certain strict arrow-of-time assumption, whose applicability may be limited to the classical domain.  This possibility has been noted repeatedly in the context of Bell's Theorem.  The argument concerning quantum computation is described here.  Further development of models which violate this strong arrow-of-time assumption, replacing it by a weaker arrow, is called for.
\end{abstract}


\maketitle



\subsection{Introduction}

Physics faces unresolved difficulties with arrows of time.  This has been evident at least since the discussions of Boltzmann's H-Theorem and Loschmidt's paradox in the late 19th century.  Although progress has been made in connecting different arrows of time to the low-entropy big-bang origin of the universe, the resulting understanding is still incomplete [see, \emph{e.g.}, \textcite{schulman1997}].  Nevertheless, ``the'' arrow of time is often taken for granted, and is familiar from the ``Newtonian schema'' of kinematics plus dynamics \cite{wharton2015a}: it is often assumed that a physical system can always be described as having a ``state'' (kinematics) which ``evolves'' (dynamics) from the past to the future.

There are also some well-known exceptions---not all physics models conform to the rules of this schema.  For example, in order to find the ``state'' of a system at a certain time according to the stationary-action principle, one must specify inputs---the values of the position coordinates---at both its past and future boundaries.  This demonstrates the ``Lagrangian schema,'' which requires an all-at-once or block-universe approach.  By looking beyond the standard schema, one is freed from the limitations of conventional thinking, and is open to novel possibilities.  Seeking such freedom is especially relevant when an impasse is encountered; this article sets forth the claim that the surprising power of quantum computing [\emph{i.e.}, its tension with the strong form of the Church-Turing Thesis \cite{arora2009}] is just the type of ``paradox'' which calls for abandoning the standard arrow of time.

There already exist several lines of evidence that quantum physics is at issue with the standard arrow of time [see also \textcite{dirac1938,wheeler1945,wheeler1949} in the classical context].  Early examples include discussions of the Einstein-Podolsky-Rosen (EPR) ``paradox'' \cite{costa1953} and delayed-choice experiments \cite{bohr1935,wheeler1978}; recent examples include argumentation from time symmetry \cite{leifer2017a} and the Pusey-Barrett-Rudolph (PBR) theorem \cite{pusey2012} [the latter enquires whether the quantum state is ontic or epistemic, \emph{i.e.}, whether it describes reality or merely the information one has regarding a system; see, \emph{e.g.}, \textcite{leifer2014}].  The strongest argument involves Bell's Theorem [see, \emph{e.g.}, \textcite{price1997}].

As typically understood, Bell's theorem proves that there is no hope for reformulating Quantum Mechanics (QM) in terms of ``hidden variables,'' or parameters directly describing events in spacetime.  But this relies on accepting the standard arrow-of-time rule, which is taken for granted by the ``Local Causality'' (a.k.a.\ ``Einstein Locality'') assumption of Bell's Theorem.  Within such an approach, QM typically describes the ``state'' of a many-particle system as a ray in an abstract and exponentially large Hilbert space, with typical applications involving superpositions and complex probability amplitudes.

Considering an alternative schema opens up the possibility of describing quantum entanglement in terms of spacetime-based parameters with standard probability rules \cite{argaman2010,almada2016}.  The apparently nonlocal connection between distant regions $a$ and $b$ is achieved through intermediate ``hidden'' parameters $\lambda$, situated far enough in the past that they could reasonably serve as a classical common cause for events in $a$ and $b$.  In defiance of the standard arrow-of-time rule, $\lambda$ is taken to depend on the inputs in $a$ and $b$, which are thus indirectly connected, through the overlap of their past lightcones.  The fact that the ``hidden'' $\lambda$ (a microscopic parameter) may depend on future inputs need not lead to violations of Signal Causality, just like the collapse of the wavefunction at $b$ due to a measurement at $a$ does not lead to violations of Signal Locality in the standard discussion of Bell correlations.  For this reason, this type of ``retrocausality'' cannot lead to paradoxical causal loops.  [Any attempt to ``measure'' $\lambda$ so that its value will be correlated with that of a macroscopic pointer would result in loss of the entanglement, as in a ``which path'' detection in the context of two-slit interference; see, \emph{e.g.}, \cite{buks1998}.]

So far, progress in developing a full reformulation of QM along these lines has been slow [see \textcite{wharton2019} for a recent review].  If too much freedom is allowed, one might obtain models with backward-in-time signaling, and it has been argued that preventing this requires fine tuning \cite{wood2015}.  Although counterarguments are available \cite{almada2016}, it seems that a physical principle, perhaps associated with the entropic arrow of time, is needed.  Such a principle could limit the excess freedom resulting from removal of the standard arrow-of-time condition, and lead to results which would systematically conform to the Signal-Causality arrow-of-time rule.

A closely-related issue has to do with the degree of correlations allowed in classical, quantum, or general non-local theories.  In proving Bell's Theorem, one typically derives the CHSH inequality --- the fact that in any locally causal mathematical model, a certain combination of correlators cannot possess a value larger than $2$ \cite{clauser1969,clauser1974}.  In general, one can generate models where this combination achieves values up to $4$ \cite{popescu1994}, but in QM its value is limited to $2\sqrt{2}$ \cite{cirelson1980}.  Again, it appears that a physical principle is involved in limiting the exaggerated freedom of generic models.  In fact, research in this context has already made significant strides, involving several suggested principles \cite{linden2007,pawlowski2009,navascues2010}.

In the present work, it is suggested that the algorithmic complexity achievable with quantum computation similarly provides motivation for rejecting the standard arrow of time.%
\footnote{Other approaches connecting the flow of time with quantum computation can be found in \textcite{aaronson2005} and \textcite{castagnoli2019}.}
Furthermore, here too it appears that a physical principle remains to be identified, one that would limit the freedom obtained with such a rejection.  The argument is based on the distinction between a Directed Acyclic Graph (DAG) and a non-directed graph.

Describing natural laws in spacetime in terms of mathematical parameters, and discretizing spacetime into $N$ distinct events, leads to a DAG if the strong arrow of time is maintained.  Assuming that the laws are local, and that the past is fixed and the future is not yet relevant, the mathematical rules for each event are greatly simplified, and the number of steps in a simulation of the physics is just the number of events, $N$.  But it is not clear to begin with that the arrow of time must be imposed [see, \emph{e.g.}, \textcite{wharton2015a}].  In particular, if there are stochastic rules that determine only how the probabilities for each event depend on events in its vicinity (in both space and time), without imposed arrows, finding the overall distribution for $N$ events may be a much more complicated computational task, due to the requirement that all $N$ events ``simultaneously'' conform to the physical laws.  As an example, consider the task of finding the ground state of a three- (or higher-)dimensional spin glass, which is known to be an NP-complete problem \cite{bachas1984,barahona1982}.

It is thus seen that if one assumes that Nature supplies us with finite ``machines'' which operate according to local rules subject to the strong arrow of time, all that can be achieved algorithmically is similar to a standard algorithm with $N$ steps, taking $N$ to appropriately represent the finiteness and the resolution pertaining to these ``machines.''  However, if the ``machines'' provided by nature are not subject to an arrow-of-time rule, the possibility that they might be capable of performing exponentially harder tasks appears to be open.  Our best understanding of quantum computing does not lead us to expect natural ``machines'' to be able to solve NP-complete tasks.  The complexity class associated with quantum ``machines'' is the BQP class, which is (to the best of our knowledge%
\footnote{Strictly speaking, that BQP is weaker than NP but stronger than P is not a proven fact, but a conjecture which is assumed here.
}) much weaker \cite{nielsen2000}.  Thus, again, a physical principle which is weaker than the standard arrow of time is required, one that would limit the achievable complexity class from NP-complete to BQP (not to P).

In the next two sections we will go through the above argument in a little more detail, describing the connection between events in spacetime and algorithmic steps in the presence (Section B) and in the absence (Section C) of a strong arrow of time.  A discussion section and conclusions follow.

\subsection{The strict arrow of time motivates the strong form of the Church-Turing Thesis}

Mathematical models of classical physics employ local variables or parameters with a clear association of a place and a time for each parameter.  A typical example is provided by the values of the classical electric field, $\bm{E}(\bm{x},t)$.  In order to connect this with algorithmic complexity, it is appropriate to discretize spacetime, taking a finite number $N$ of events, $(\bm{x}_n,t_n)$, distributed reasonably uniformly, to provide a sufficiently detailed representation of a finite region of (Minkowski) spacetime, to some desired accuracy.

Within a kinematics plus dynamics schema, the state of the modelled system at time $t$ would be represented in this picture by the events $m$ with times $t_m$ between $t-\Delta t$ and $t$ for an appropriate small $\Delta t$, and the values of the model parameters $\mu_m$ associated with these events.  The model obeys Local Causality if the dynamics specify a rule (which may be either deterministic or probabilistic) for obtaining the value of the parameters at the $n$th event from the parameters in its recent past and its close vicinity, with spacelike separations avoided, so that the relevant events are in the past relative to $t_n$ in all frames.  We will denote the set of indices of these earlier and nearby events by $r(n)$.

If an external input, such as an external force, acts at the $n$th event, the value of the parameters at that event will be affected, but the values at earlier times will not.  The parameters $\mu_n$ associated with the $n$th event thus include inputs $I_n$ and non-input parameters $Q_n$ (each of these is in general a set of parameters, not limited to scalars).  In the deterministic case, the dynamical rule $F_n$ specifies the value of $Q_n$ as a function of $I_n$ and the earlier $\{\mu_m\}_{m \in r(n)}$ (the rule $F_n$ depends also on the spacetime locations of $n$ and the $m$s, of course).  For example, a model discretizing Maxwell's equations in this manner would have $Q_n$ corresponding to the electromagnetic fields, and $I_n$ specifying the charge and current densities, the relevant inputs in this case.  (For stochastic models, $F_n$ determines the probability distribution of $Q_n$.)

Assuming that the values of the parameters $\mu_n$ are appropriately discretized as well, so that the modeling of each of the $N$ events is finite, this description makes it obvious that the algorithmic complexity of a simulation according to such a model is $O(N)$.  Conversely, the modeled physics cannot provide results which are not efficiently achievable by an algorithm with $O(N)$ steps.  That physical systems are (polynomially) equivalent to algorithms in this sense is an expression of the strong form of the Church-Turing thesis \cite{arora2009}.  Barring problems with the discretization scheme, classical physics indeed operates in this manner.  That quantum physics is different is discussed in the next sections.

Note that we have here considered only the number of steps in the algorithm, $N$.  It is of course possible for only $M$ out of the $N$ events to have external inputs, such as initial conditions, with the other $N-M$ events having no inputs (or having the corresponding $I_n$ set to zero or null in some fashion).  It is further possible to have the number of physical parameters $N$ exponentially larger than the number of physical inputs $M$, but this possibility is not of interest for the purposes of the present discussion, which focuses on $N$ itself.

It is natural to take the $N$ spacetime events to be nodes of a graph, with directed edges from the $m$s in $r(n)$ to $n$ itself, representing the dynamical rules $F_n$.  The resulting DAG (Directed Acyclic Graph) represents the discretized mathematical model, as well as the algorithm which would carry out a simulation according to the model.%
\footnote{This $O(N)$ graph is not to be confused with the exponentially large configuration graph representing all the possibilities for a model \cite{arora2009}.}
All the edges in the graph are directed from the past to the future (that the graph is acyclic corresponds to assuming a standard Minkowski geometry, with no closed time-like curves).

Note how central the standard arrow-of-time assumption is to the logic leading to the Church-Turing Thesis (it is not denied, of course, that there are additional necessary assumptions, \emph{e.g.}, regarding discretization---for the present purposes it suffices that the standard arrow of time is one of the necessary assumptions).  It is the assumption that for each node $n$, the task of obtaining the parameters $Q_n$ can be performed while taking all the parameters from the past, \emph{i.e.}, from $r(n)$, to be fixed, and ignoring all the parameters relating to the future, which leads to the finiteness of this task.  Performing this for $N$ nodes is then necessarily an $O(N)$ task.

\subsection{Models with no arrow of time}

Removal of the above arrow-of-time restriction would have dramatic consequences for the algorithmic-complexity consideration.  If a mathematical model is associated not with a DAG but with an undirected graph of $N$ nodes, its capacity for computation could be entirely different.  In fact, for a reasonable choice of the replacement for the rules $F_n$, the category of computations which can be efficiently performed by a ``machine'' which would efficiently generate solutions of the relevant model would be the NP class.  As already mentioned, this is known if the rules are replaced by those of a spin-glass system \cite{bachas1984,barahona1982}.  (Which rules should be expected for future physics theories is of course completely open---for example, at the dawn of QM, Heisenberg employed noncommuting-operator rules while Schroedinger used differential equations.)  A further simple example is described next.

The example involves a standard NP-complete problem, such as scheduling $M$ meetings within a finite given time $T$.  The requirements concerning the length of the meetings and the intended participants are to be specified by inputs $I_m$, and the timing of each meeting by parameters of a different type, $P_m$.  As the problem is NP, it is known that $N$ steps are required to verify that the meetings have no conflicts, with $N$ polynomial in $M$.  It is easy to construct a DAG with $N$ nodes representing the algorithm for performing this verification process, beginning with the $I_m$s and $P_m$s as inputs and resulting in an output $O$ which is true for a valid combination of the timings $P_m$.  The directions of the edges of the DAG lead from its inputs $I_m$ and $P_m$ to its output $O$.  Each of the $N$ steps is associated with a rule $F_n$, consistent with the description of the previous section.

Consider now removing the arrows from the graph.  This could represent a model where each rule $F_n$ is replaced by a weight $W_n$, which depends on the same parameters involved in $F_n$.  Thus, the rules of the model are local as before, but the model dictates the overall behavior of the combination of parameters $\{ Q_n \}$, and cannot be easily separated into $N$ consecutive steps.  Combining all of the local rules involves multiplying the weights $W_n$ for all $n$, and normalizing the weights to obtain a probability distribution involves adding all the product weights, resulting in a normalizing factor $Z=\sum_{ \{ Q_n \} } \left( \prod_n W_n \right)$.  (In the statistical mechanics context, $Z$ is called the partition function, and the weights are given by an exponent involving the potential energy and the temperature.)

Returning to the specific scheduling problem above, one can define each of the weights $W_n$ as equal to unity for every combination which is consistent with the rule $F_n$, and to zero otherwise.  One may further set the inputs $I_m$ for a specific scheduling task, and set the ``output'' $O$ to ``true.''  If the time $T$ is not too short,%
\footnote{Dealing with shorter times, or with tasks for which the structure of the graph and the $F_n$ rules depends not only on the $I_m$s but also on the $P_m$s, requires more-complicated examples.}
every valid schedule, \emph{i.e.}, every valid combination of the $P_m$s, together with the corresponding values of the other $Q_n$ parameters, would have a weight of unity, and all other combinations would have a vanishing weight ($Z$ represents the number of valid schedules).  The result of such a model would be to generate at random one of the valid schedules.  This too is, of course, an NP-complete task.

The resulting pattern is similar to that described for the CHSH inequality in the introduction.  The standard arrow of time is a strong restriction that would limit the capabilities of any model of a physical system to those of standard algorithms, in accordance with the strong form of the Church-Turing Thesis.  Quantum systems are not as limited as that, but removing the arrow-of-time restriction altogether would result in capabilities which are ``too powerful'' according to reasonable expectations.  A restriction is necessary, but it needs to be less powerful in order to curtail the achievable complexity class from NP to BQP, not to P.

\subsection{Discussion}

The above argumentation may appear to be making a philosophical point, but the intention is to apply this reasoning to guide developments in mathematical physics.  This follows \textcite{feynman1964}, who contended that nobody understands QM.  He demonstrated how predictive power can coexist with a lack of understanding with the example of the astronomers of the Maya culture, who possessed a mathematical procedure for predicting the appearances of the moon---the timing of a new moon or of an eclipse---which did not involve any conception of orbital paths.  Feynman also suggested that developing reformulations of existing theories can serve to improve our understanding, even if no novel predictions are involved (examples include the development of Lagrangean and Hamiltonian mechanics as alternative formulations of Newton's equations; for a long time, these only improved our understanding of classical mechanics; much later, they also played essential roles in the development of QM).

In this context, the upshot of the previous sections is that quantum computation adds to our motivation to develop reformulations of QM which do not conform to the standard arrow of time.  But it is clear that some effective arrow of time must be retained.  Physical theories in general, and standard QM in particular, conform to the Signal-Causality rule---they describe signaling to the future, but not to the past (there are many aspects which are time reversal-symmetric, but there is always something to break the symmetry, often just a special treatment of initial conditions).  Thus, the flow of accessible information, relating to the inputs and the outputs of the theory, is always from the past to the future.

In standard Schroedinger-picture QM, this past-to-future flow affects the internal parameters of the theory as well---the quantum state or wavefunction is taken to evolve from the past to the future (whether or not collapse is allowed for).  A reformulation breaking the standard-arrow-of-time rules would involve some internal parameters which depend on other parameters in their future (possibly a statistical dependence, \emph{i.e.}, having a probability distribution which depends on future parameters).  In order for this future-dependence to play an essential role, it must involve relationships which cannot be simply inverted, such as a dependence on the externally-controlled settings of future measurement devices.  For this reason, the relevant arrow-of-time condition involves future input parameters, and is called No Future-Input Dependence in \textcite{wharton2019}.

The situation concerning causality or the arrow of time in reformulations of QM which would violate this condition is similar to that concerning locality in standard QM, which violates locality in the sense of Bell's Local-Causality condition, but conforms to Signal Locality.  Here the No Future-Input Dependence condition would be violated for internal parameters, but the output parameters would not have this characteristic---the Signal Causality condition involving the outputs would be maintained.

As noted in the introduction, relaxing No Future-Input Dependence has dramatic consequences for reformulations of QM.  The generalization of Bell's locality condition to models with Future-Input Dependence is called Continuous Action, and maintaining this locality condition has distinct advantages, in addition to the necessary Signal Locality \cite{wharton2019}.  In fact, Bell's Local Causality condition can be seen to follow from requiring both Continuous Action and No Future-Input Dependence (assuming Lorentz Covariance and the use of standard mathematics and probability rules).  Thus, if a reformulation of QM with Continuous Action can indeed be found, it will accordingly be based on a model with parameters with Future-Input Dependence.  

It would be natural to view these parameters as providing a more-or-less direct description of reality --- ontic variables --- with the standard ``quantum state'' taken to merely represent the information available to an external observer up to a time $t$.  This is the psi-epistemic view of QM [see, \emph{e.g.}, \textcite{caves2002}].  The arguments posited against this view in the past would fail in the presence of Future-Input Dependence.  The fact that this state ``evolves'' with $t$ in an information-conserving manner (unitarity) would be required by its role as representing unchanged information, as long as indeed there is no update of the available information.  Similarly, this ``state of knowledge'' would have to suddenly change upon such an update, explaining precisely why and how measurements cause ``wavefunction collapse.''

This brief discussion only aims to indicate that the development of Future-Input Dependent models with Continuous Action is feasible in principle.  For details, including concrete examples of toy models reproducing QM in the specific context of Bell's Theorem, see \textcite{wharton2019}.  Developing a full reformulation of QM along these lines appears to be challenging not because of a necessity to deal with a particularly complicated situation, but primarily because of the need to overcome the barrier associated with conventional thinking concerning the arrow of time.

\subsection{Conclusion}

When we use a mathematical model to describe the objective properties of a physical system, we generally expect these properties to depend on the past of the system, not on its future.  This works well in the classical, macroscopic domain, but the presence of quantum fluctuations and uncertainty appear to undermine such thinking for quantum systems.  The time-symmetry of microscopic physical laws similarly speaks against such a distinction between the past and the future.%
\footnote{Indeed, such time-symmetry arguments have been used to motivate alternative time-symmetric interpretations of Quantum Mechanics, both the transactional interpretation \cite{cramer1980} and the two-state-vector formalism \cite{aharonov1991}, but these approaches still employ the standard quantum ``state,'' which for many-particle systems is exponentially complex and cannot be represented in terms of local variables $\mu_n$.}
Allowing the system's ``objective'' microscopic parameters to depend on the specification of the measurement to be made on the system at a later time, not only on the earlier preparation, may resolve many a quantum mystery.  As described above, the ``nonlocality'' of Bell's Theorem serves as the prime example --- quantum phenomena violate the relevant ``no-action-at-a-distance'' condition only when this condition is formulated within models with such a strong past-future distinction.  

Generalizing the ``no-action-at-a-distance'' condition to models which are time-reversal symmetric, or which possess a weaker arrow-of-time rule, removes the restriction posed by Bell's Theorem \cite{wharton2019}.  This could serve to ``explain'' the power of quantum computation --- if indeed microscopic parameters are not subject to the rules of a DAG, the associated complexity class need not be limited to P.

Once this point of view is accepted, one is faced with a sharply contrasting problem.  It is not that quantum computation is surprisingly powerful --- it becomes surprising that it is not even more powerful.  A ``physical principle'' must be imposed on the relevant family of models to limit the capacity from NP to BQP.  This is closely analogous to the search for a limiting physical principle in the context of Tsirelson's bound, which is related to Bell's Theorem and has been an active field in recent decades.  Perhaps concepts from quantum computation will provide additional clues or lead to new directions on this adventure.

Acknowledgements: The author wishes to thank Scott Aaronson and the other participants of the 6th FQXI conference (Castelvecchio Pascoli, Italy, July 2019) for thought-provoking discussions, and Oded Schwartz and Ken Wharton for helpful comments on a draft of the manuscript.

\bibliography{QCandAoT}

\end{document}